# An Interpretation of Milne Cosmology

Alasdair Macleod
University of the Highlands and Islands
Lews Castle College
Stornoway
Isle of Lewis
HS2 0XR
UK
Alasdair.Macleod@lews.uhi.ac.uk

*Abstract* The cosmological concordance model is consistent with all available observational data, including the apparent distance and redshift relationship for distant supernovae, but it is curious how the Milne cosmological model is able to make predictions that are similar to this preferred General Relativistic model. Milne's cosmological model is based solely on Special Relativity and presumes a completely incompatible redshift mechanism; how then can the predictions be even remotely close to observational data? The puzzle is usually resolved by subsuming the Milne Cosmological model into General Relativistic cosmology as the special case of an empty Universe. This explanation may have to be reassessed with the finding that spacetime is approximately flat because of inflation, whereupon the projection of cosmological events onto the observer's Minkowski spacetime must always be kinematically consistent with Special Relativity, although the specific dynamics of the underlying General Relativistic model can give rise to virtual forces in order to maintain consistency between the observation and model frames.

## I. INTRODUCTION

Edwin Hubble's discovery in the 1920's that light from extra-galactic nebulae is redshifted in linear proportion to apparent distance was quickly associated with General Relativity (GR), and explained by the model of a closed finite Universe curved under its own gravity and expanding at a rate constrained by the enclosed mass. The model prevails to this day. At the time, not everyone agreed with this interpretation. Edward Arthur Milne argued that curved spacetime and suchlike were unobservable and a consistent model should be presented only in terms of observables. This so-called operational approach presumed spacetime to be flat (because it appeared so) and therefore all observations should be subject to Special Relativity (SR). The redshift of distant nebulae is then required to be a Doppler effect associated with a proper recessional velocity [1]. By rigorously applying SR, the apparent distance-redshift ($D_L$-$z$) relationship that emerges from the analysis is in surprisingly good agreement (though not perfectly so) with the current observation of distant 'standard candles' and is completely consistent with other cosmological data. Of course, standard GR cosmology also predicts the apparent distance-redshift relationship for standard candles. By comparing with supernova Ia observations, the favoured ΛCDM model is found to be a better match for the data than Milne's model. Nevertheless, it is curious how two apparently unrelated approaches to cosmology can lead to very similar predictions as far back as 8 billion years. The normal explanation is to claim that the Milne procedure is just another derivation of a very specific GR model, an empty Universe, to which the real Universe approximates. This is very reasonable, of course, but requires the existence of an appropriate coordinate change that will transform the cosmological redshift into a Doppler shift.

There is an alternative explanation that emerges naturally from the recent discovery that the Universe is flat [2]. From this finding, it follows that our observation platform as we look out upon the galaxies is Minkowski spacetime and that the elegant underlying General Relativistic model is not necessarily directly apparent to us. The General Relativistic model may explain the expansion of the Universe and drive the dynamics, but cosmological events are observed in (and are a projection onto) flat spacetime with kinematics in this frame subject to the rules of SR, rules that give rise to the equations of Milne Cosmology. The similarity between Milne Cosmology and the concordance model is then not mysterious but merely an observational effect that requires no further explanation. The key point in this argument is that a clear distinction must be made between models, which purport to explain structure and causes (the 'Why?'), and observational frames which simply impart consistency and causality on observation (the 'How?').

However, it can also be argued this approach does not actually explain the similarity in the predictive power of Milne Cosmology and the concordance model, but merely reintroduces the puzzle in a different guise. As we will see, a key feature of Milne Cosmology is that apparent recessional velocities remain constant over time from energy conservation considerations; why then does the General Relativistic model when projected onto flat spacetime show cosmological entities with a largely constant expansion velocity when there is no obvious principle which makes this a necessity? This important point will be considered in the final section.

## II. COSMOLOGICAL EVENTS

Consider a stationary observer $O$ in Minkowski spacetime observing events $\varepsilon_1$ and $\varepsilon_2$ with coordinates $(x_1, t_1)$ and $(x_2, t_2)$ respectively. A moving observer $O'$ co-located with $O$ at $t = 0$, will perceive the same events to have originated at coordinates $(x_1', t_1')$ and $(x_2', t_2')$ where the relationship between the primed and non-primed coordinate sets specifies the Lorentz transformation rules. This simple example shows the positioning and timing of events is flexible and alters in such a way that the conservation laws always hold and the speed of light is the same in any inertial frame. Consider now the possibility that these are actually cosmological events (events originating from bodies subject to the cosmological expansion with respect to the observer). Are they distinguishable from non-cosmological (local) events by any observation procedure? We may introduce local events with identical 4-coordinates to shadow cosmological events. Whether this action is physically possible to implement is another matter, nevertheless one can conceive of such an experiment. In this situation, we are really asking if there exists a distinction in the relationship between events within the local and cosmological event sets. What is absolutely certain is that any difference between the event sets cannot be explained by the introduction of a simple force, field or geometrical effect, but instead implies a violation of energy conservation and/or a variation in the speed of light, both of which are disagreeable outcomes. There really is no element of freedom in SR as it currently stands that can be utilized to deal with spacetime expansion and its effects. If SR is not to be reworked, it is necessary that events on the expanding spacetime manifold should be replaceable with equivalent local events that are fully consistent with SR constraints. We therefore conclude that the cosmological expansion can be treated as an instantaneous proper velocity when viewed in Minkowski spacetime.

Fig. 1 illustrates a generalised form of the relationship between frames.

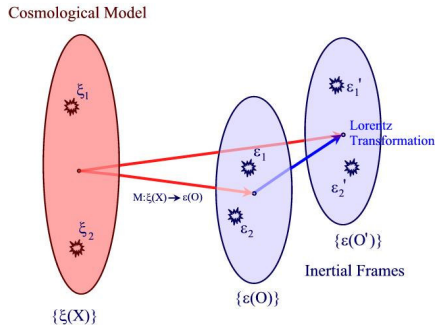

**Figure 1** Set representation of events in cosmological coordinates observed from two distinct inertial frames. The mapping *M* between the domain and one codomain must be injective and surjective if the detail of the cosmological model is to be extracted by correlating observations from an inertial frame.

What is uncertain is how cosmological events are mapped onto the inertial frames (the mapping *M* in the diagram). Certainly we do not have the right to expect a relation as simple as the Lorentz transformation between inertial frames. Consequently it is unclear if the complete detail of the underlying cosmological model can ever be deduced by piecing together observational information in the Minkowski frames. An obstacle here is that SR is an observation schema and is unconcerned with models. We can see this in a non-cosmological context by considering a group of inertial observers with different relative velocities. It is not possible to reconcile their instantaneous worldviews into a single absolute spacetime arrangement of entities that all observers will find consistent. At the very least, local clocks will drift.

Whilst the detail of the cosmological model driving the dynamics is unimportant, the mapping of events must nevertheless conserve energy. The energy loss in flat spacetime associated with the apparent recession velocity will be a Doppler effect (again, the constancy of the speed of light and energy conservation compel this). It is unnecessary (and unlikely) for this to exactly match the cosmological energy loss resulting from the expansion of spacetime as the photon propagates. Any discrepancy can be compensated for with the appearance of an apparent acceleration in the Minkowski frame. The observation of events tied into the expanding spacetime manifold may therefore give rise to virtual forces that have their origin in the structure of the expanding Universe. These forces distinguish between entities that are part of the expansion and those that are not.

We can conclude from this section that it is impossible to construct cosmological models by merely applying SR, yet this was exactly what Milne did. How did he construct an apparently effective model, at least in its predictions, in spite of these caveats? To answer this question, it is necessary to carefully consider his methodology.

## III. PRINCIPLES OF MILNE'S COSMOLOGY

Before one can objectively evaluate the SR procedure used by Milne, it is desirable to separate the mathematical method from the cosmological model used in its derivation. Mathematically, the analysis proceeds by specifying initial conditions whereupon the kinematic development over time is obtained by applying the axioms of SR and the conservation laws. The initial (or boundary) conditions are unusual. Essentially, all the matter in the Universe is thought to be accumulated at a single spatial point at $T = 0$. This is also the synchronisation point for all clocks. It is uncertain what the state of matter should be under this infinitely dense condition, but is in many ways irrelevant as we are not concerned with issues such as the subsequent growth of structure. To present the argument without unnecessary complication, a simple aggregation of non-interacting particles will therefore be assumed. In addition, each particle is allocated a random velocity in the range 0 to *c*. Because spacetime is flat, there can be no acceleration associated with geometric effects. Therefore, in the absence of forces, each particle maintains its initial velocity as the system develops over time. Thus the particles move apart and separate into expanding shells with radial velocity proportional to the shell radius.

Milne also understood the role of SR as simply 'a technique, a calculus' [3] and recognised that SR in itself cannot be used to build models – models must be constructed around the formalism and do not necessarily emerge from it. Without a model, one cannot say what all the matter, initially accumulated at a single point, is moving into. It does not matter whether the space already exists (as Milne thought), is being continuously created (GR big bang cosmology), or does not really exist at all (the relationist view); the procedure being applied is unaffected. Nor do we concern ourselves with why the initial velocities arise, why the mass is accumulated at $T=0$, or why particles are subsequently free to move.

We will show in the next section that the characteristic equations of Milne Cosmology can be derived by the application of SR to some very specific boundary conditions. We are essentially presenting SR cosmology stripped of its philosophical baggage to expose the mechanism that leads to the effective $D_L$-$z$ relation. Milne, in contrast, developed his theory working on the principle that one can 'attempt a complete reconstruction of physics from the bottom up, on an axiomatic basis' [3]. Milne attached a physical interpretation to both the initial conditions and the form of spacetime. This resulted in additional predictions that were contradicted by later observational data. Nevertheless, the model conceived by Milne is highly imaginative. Milne envisaged a finite bounded Universe that could be made to hold an infinite quantity of matter by a neat trick. He recognised that greater and greater quantities of matter can be packed into space with increasing recessional velocity by exploiting the Lorentz contraction effect. The complete development of the model [1] gives rise to two time scales, a possible variation in the gravitational constant with time and a particular redshift-luminous mass density relationship, none of which have been verified. The model that Milne attached to his procedure has therefore been convincingly refuted; however the basic SR analysis has not. By rejecting Milne's interpretation but retaining the methodology, we are separating the discredited model from the analysis. Essentially, momentum conservation and some very specific initial conditions are sufficient to develop a $D_L$-$z$ function that can be compared to observational data. The Hubble relationship naturally emerges as a consequence of the analysis.

It is possible to argue that the initial conditions chosen by Milne may be unrealistic and are rather contrived; but when we explore the effect of small changes to these conditions, we find that the equations are quite resilient, and resistant to change. In fact, the key assumption is found to be that the expansion velocity is maintained over time because there is no nett gravitational effect. We can immediately see the connection to the General Relativistic empty Universe model which shares the property that there is no constraint on the expansion. However, it is another matter to claim the SR model is equivalent to an empty Universe: This particular solution to Friedmann's equation describes a linear coasting Universe [4] expanding at a constant rate $c$ and is frequently used as a reference against which other cosmologies are evaluated. There is a general belief that the SR spacetime and the empty Universe are topologically identical and even isomorphic. Thus the Milne model is very persistent because of its migration into GR cosmology and is often referred to as a toy Universe model and oft quoted for instructive reasons. Although the models can be topologically matched by a suitable coordinate change (leading again to two distinct time scales), it is unclear how the incompatible SR and GR definitions of redshift can ever converge (see *Appendix 1* for a more quantative discussion). Even if this difference could be reconciled, it is troubling that Milne Cosmology maps onto such an unrealistic GR model – the Universe is clearly not empty.

## IV. THE SR DISTANCE-REDSHIFT RELATION

The relationship between the apparent cosmological distance (as determined from the received energy flux) and redshift in the observer Lorentz frame is often derived by stating the Hubble law and replacing the velocity term in the Hubble law with the Doppler redshift expression. See, for example, Davis and Lineweaver [7]. The problem is that the correct form of the Hubble law in the context of SR is not known *a priori*, hence derivations of this type are unsatisfactory. The expansion velocity can be related to the time at photon emission, the time at photon absorption, the photon travel time, the proper distance at emission, the proper distance at absorption, the retarded distance and so on. The expression may even be a combination of these terms – there is no certain indication as to the true expression. If instead the initial conditions are known, the Hubble law emerges from an analysis of the development of the system over time. This is therefore the preferred approach, and although it may seem at first glance there is a strong dependence on what can only be described as arbitrary initial conditions, we will show this is not actually the case.

We will therefore begin with the Milne boundary condition where all mass is accumulated at $T=0$ with a distribution of particle velocities ranging from 0 to $c$. Chodorowski [8] derived the correct $D_L$-$z$ relationship starting from the same point and following a similar procedure. The derivation here is slightly different but gives rise to the same final expression. Our emphasis is on the strong connection between the initial or boundary condition and the form of the Hubble relationship that emerges. This in turn impacts on the apparent distance-redshift function. We then have a means of exploring how the initial condition affects the $D_L$-$z$ function.

We can derive the distance redshift relation without loss of generality by considering two particles moving apart at a velocity $v$ starting out from $T = 0$ when their positions coincide (Fig. 2). In the observer Lorentz frame, we can trace back photons received at $T = T$ from an event originating in the moving frame. The original event has local coordinates in the moving frame of $(0, \gamma[1-\beta]T)$, where $\beta = v/c$ and $\gamma = (1-\beta^2)^{-1/2}$. The event first appears in the observer frame with local coordinates $(vT/[1+\beta], T/[1+\beta])$. Photons associated with the event then propagate to the stationary observer and are registered at time $T$.

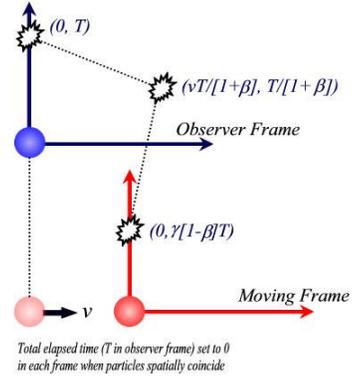

**Figure 2** An event in the moving frame is perceived in the observer frame as shown [9].

Of interest is the measured energy flux compared to a stationary source. Photon effects reduce the energy density by a factor $(1+z)^2$, where $(1+z) = (1+\beta)^{1/2}/(1-\beta)^{1/2}$. This is because the photon oscillation frequency is reduced by $(1+z)$ because an interval $T$ in the observer maps onto $T/(1+z)$ in the emission frame. There is a second factor arising from a difference in the rate at which photons are emitted and received; basically, the rate is reduced by a time dilation factor $\gamma$, and because the space between is the emitter and absorber is growing, and the growing space becomes filled with in-transit photons, there is an associated attenuation factor of $(1+\beta)$. Together these two elements make up the second power of $(1+z)$. It may appear double counting taking place here, with adjustments for both the wave and particle aspects of the photon, but this is permissible and necessary.

There is another source of flux diminution associated with the spread of energy over a larger spherical surface at absorption compared to the situation with the emitter and absorber at rest with respect to one another. This flux diminution is proportional to the square of the proper separation at emission in the Lorentz frame of the observer, multiplied by a factor $(1+z)^2$. (The theoretical derivation is given in Misner, Thorne and Wheeler [10], sections 22.6 and 29.4.) From Fig. 2, the proper distance at emission in the Lorentz frame of the absorber is $vT/[1+\beta]$ (actually the photon travel distance). The relationship between flux ($f$) and bolometric luminosity ($L$), after accounting for each effect is

$$f = \frac{L}{4\pi D_e^2 (1+z)^4}, \quad (1)$$

where $D_e$ is the proper separation at emission measured in the observer frame ($T$ = constant). The veracity of the equation is demonstrated by the fact that the CMBR, which is emitted as a blackbody spectrum, is still observed as a black body spectrum. This is only possible with a $(1+z)^4$ diminution in flux with redshift. Equation (1) can also be derived by imposing energy conservation on the normal Doppler effect subject to the condition that the relative velocity remains constant for the period between emission and absorption.

We can therefore define the luminosity distance $D_L$ as

$$D_L = (1+z)^2 D_e = (1+z)^2 \frac{vT}{1+\beta} \quad (2)$$

because $D_e$ is the velocity times $T/(1+\beta)$. This is the same expression obtained by Chodorowski [8].

By relating the velocity to the redshift using the SR Doppler relationship $(1+z) = (1+\beta)^{1/2}/(1-\beta)^{1/2}$, we can present the expression in terms of direct observables (since the luminosity distance can be derived from the apparent magnitude of standard candles):

$$\frac{D_L}{cT} = \frac{H_o D_L}{c} = z + \frac{z^2}{2}. \quad (3)$$

Since we can only really only make measurements when $T = T_o$, we have taken the liberty of substituting the value of the Hubble constant in the present epoch, $H_o$, which is defined as $1/T_o$ in SR because the Milne Universe expands in at a constant rate $c$. In any case, it is conventional in standard GR cosmology to identify $H_o$ with the inverse of the current age of the Universe. The Hubble Law emerges from this model and takes the explicit form

$$D_e = \frac{vT_o}{1+\beta} = vT_e \quad (4)$$

Equation (4) is quite different from the normal statement of the Hubble law: $v = cT_\gamma/T_o$, where $T_\gamma$ is the photon travel time. Equation (4) in terms of these parameters is $v = (1+\beta)cT_\gamma/T_o$. The equation can also be expressed in terms of the current value of the Hubble constant:

$$v_H = \frac{H_o D_e}{1 - \frac{H_o D_e}{c}}. \quad (5)$$

The reason for the difference in the $D_L$-$z$ relationship derived by Davis and Lineweaver [7] and Chodorowski [8] is now clear – two completely different forms of the Hubble law are used. The Milne form (equation (5)) is clearly preferable in that it emerges from the analysis, but how much is it a feature of the initial conditions? We will consider this in the next section.

As an aside, it is now easy to show that Milne cosmology makes equivalent predictions to the empty Universe in General Relativistic cosmology. In GR, the relationship between $z$ and the luminosity distance is [2]

$$D_L = \frac{c}{H_o}\left\{z + z^2\left(\frac{1-q_o}{2}\right) + \sum_{i=3}^{\infty} f_i(z^i)\right\}. \quad (6)$$

The equation is approximated by neglecting the higher order terms and expressing the deceleration parameter $q_o$ in terms of the matter and cosmological parameters, $\Omega_M$ and $\Omega_\Lambda$:

$$\frac{H_o D_L}{c} \cong z + \frac{z^2}{2}(1 - \frac{\Omega_M}{2} + \Omega_\Lambda). \quad (7)$$

If $\Omega_M$ and $\Omega_\Lambda$ are set to 0, the equation is the same as the SR $D_L$-$z$ relationship, equation (3). However, this is not the only equivalence. If $\Omega_M = 0.66$ and $\Omega_\Lambda = 0.33$, the models again converge. The current evidence is that $\Omega_M = 0.27$ and $\Omega_\Lambda = 0.73$, the concordance values. The Milne $D_L$-$z$ function therefore deviates from the concordance model as $z$ approaches and exceeds one in a small but significant manner.

## V. THE MILNE BOUNDARY CONDITIONS

By adopting a specific boundary condition, a particular form of the Hubble law emerged which determined the $D_L$-$z$ function. Although similar to the corresponding $\Lambda$CDM $D_L$-$z$ function, the two functions deviate significantly. We will consider if it is possible to modify the function by altering the boundary condition. Basically we have at $T=0$ all the mass in the Universe with a random distribution of velocities from 0 to $c$. Is there any evidence to support this assumption? One might argue that the system should have initially been in thermal equilibrium whereupon we would expect luminous matter when currently observed as a function of redshift to reflect the appropriate statistical distribution at $T = 0$. If this were true, there would now appear be a mass peak at the recession velocity corresponding to the blackbody temperature. This is discussed at length by Milne [1], but there is no real evidence for this type of maximum in the space distribution of luminous matter. Therefore no observational justification for Milne's choice of initial condition exists, and we are at liberty to make changes.

What changes are sensible? We know there are very old stars and the particles that make up the star have probably never individually participated in the cosmic expansion. We can therefore consider a more realistic model where particles join the expansion only under certain conditions. The favourable conditions may occur as late as billions of years into the life of the Universe, or never.

We can investigate these ideas by modifying the initial conditions to incorporate a lag in matter joining the cosmic expansion. Consider a luminous mass a distance $r$ from the observer joining the cosmological flow at time $T_e$ measured in the frame of the observer. The apparent luminosity distance at time $T (= T_o)$ is

$$D_L = (1+z)^2\left(\frac{r + v(T - T_e)}{1+\beta}\right). \quad (8)$$

This is because the proper separation at emission in the Lorentz frame of the observer is now the separation when the expansion kicked in at $T_e$ ($r$ by definition), plus the Hubble velocity multiplied the time for the duration the expansion was active, $T - T_e$. Of course, this must be divided by $(1+\beta)$ to take the propagation delay in the observer frame into account.

Equation (8) is valid for $T \geq T_e + r/c$. But a real problem now presents itself. The Hubble law previously emerged from the initial conditions. By imposing a lag, there is no way of deducing the appropriate form of the Hubble law for this situation, hence it is not possible to say what recession velocity should be attributed to an entity joining the Hubble flow when $T > 0$. If we consider the mass becoming 'attached' to the expanding spacetime manifold and carried along by the expansion (as GR supposes), it is legitimate to adopt equation (4) as the appropriate Hubble velocity. Using equation (4), we find that $v = r/T_e$ and substituting into (8),

$$D_L = (1+z)^2 \frac{vT}{1+\beta}. \quad (9)$$

This is identical to (2). Particles therefore join the motion as if they had been moving at that velocity from $T=0$. This may be seen as a circular argument, and in many ways that is true as we presuppose that expanding spacetime carries along the entities. This sort of model-based assumption is inappropriate to SR analysis. Nevertheless, we find it does not really matter what recession velocity is attributed to an entity that joins the flow at a later time. The entity will drift towards the position in the flow corresponding to equation (4) by virtue of the fact that the velocity remains constant.

Thus it would appear the SR $D_L$-$z$ relation is surprisingly resistant to change and is unaffected by complex development cycles in the Universe where masses may alternate between gravitationally bound states and unbound states that are subject to the cosmological expansion. This is both reassuring and troublesome. Reassuring in that this is consistent with observation, and troublesome because it is actually very difficult to modify the SR $D_L$-$z$ relation to match observational data by making simple changes to the boundary conditions.

## VI. DISCUSSION & SUMMARY

We have considered the projection of General Relativistic cosmological models onto flat Minkowski spacetime and stated that SR is concerned with observation and is unsuitable for the creation of models for the Universe; thus it is an error to associate the Special Relativistically derived distance-redshift relation with its own distinct cosmology. In whatever way GR explains the dynamics of the expansion, the kinematics will have a valid and consistent interpretation in SR. The coinciding of the $\Lambda$CDM and the SR model predictions with observation is just a reflection of the validity of SR in its own domain, nothing else.

By this argument, we would expect the Special Relativistic distance-redshift relation to be exactly match observation. It does not, but this need not be a major concern: The boundary condition may need modification (although we saw this was quite a difficult thing to do); spacetime may not be completely flat, there may be some residual gravitation; there may be real or virtual forces to take into account once the underlying model is fully understood. One other possibly important factor that may align data with theory is the inclusion of the reaction force on radiating bodies. This is a factor in SR analysis that cannot always be ignored.

In meandering through Milne Cosmology, we have bypassed many conceptual issues and problems. The main issue is, of course, the fact that GR encompasses SR. How then can we decouple GR cosmology and SR in the manner described in the paper? The reason is because, according to GR, inflation can make spacetime virtually flat (and this has been confirmed) but still the expansion is active over the frame. Unfortunately, there is no mechanism in SR to manage this expansion. There is clearly a correspondence problem as curvature tends to zero and spacetime becomes Special Relativistic. It is this problem the paper tries to address. Of course, one might adapt SR to incorporate the expansion, but it may be that this course of action is unnecessary. It is really for the theoreticians to explain how cosmological events viewed from Minkowski spacetime will not exhibit a Doppler effect – there is no 'get out' clause at the moment. In any case, the apparent constancy of the Hubble velocity is disturbing and without explanation.

## APPENDIX 1: REDSHIFT COMPATABILITY

Form Fig. 1, it is suggested that cosmological and observational coordinates do not directly map. In trying to equate the empty Universe and Milne models, the same coordinate systems are often used. It is easy to see the problem with this. Compare the redshift ($z$) definitions in SR and GR, equations (1) and (2) respectively using the same coordinates:

$$1+z = \sqrt{\frac{1+\frac{v}{c}}{1-\frac{v}{c}}} \qquad (A1.1)$$

$$1+z = \frac{R(T_o)}{R(T_e)} \qquad (A1.2)$$

where $R(T)$ is the scale factor (= $cT$ in the empty Universe model since $H = \dot{R}/R$), $T_o$ is the current time and $T_e$ is the time of emission. For (A1.1) and (A1.2) to be equivalent, this requires a Hubble law definition in SR of:

$$\frac{v}{c} = \frac{T_o^2 - T_e^2}{T_o^2 + T_e^2} \qquad (A1.3)$$

Whilst this reduces to a linear Hubble law for $T_o - T_e$ small (= $\Delta T$)

$$\frac{v}{c} \cong \frac{\Delta T}{T_o}, \qquad (A1.4)$$

it is difficult to relate the form of (A1.3) to a recognisable underlying physical mechanism harmonious with GR. In the section IV it is shown that (A1.3) is incompatible with SR, thus the Milne and the empty Universe models are not mathematically equivalent. In other words, GR time coordinates do not map directly onto SR time coordinates.

Attempts have been made to unify the redshift definitions. Harvey *et al.* [5] have shown that the SR redshift formula can be stated as $1+z = ds'/ds$ (Weyl's definition) where $ds$ and $ds'$ are respectively the differences in proper time at the absorption and emission ends of null geodesics. Whiting [6] has demonstrated that, under certain circumstances, the cosmological redshift can be decomposed into a relativistic Doppler redshift and a gravitational redshift. In spite of this progress, the situation is still problematic.